\documentclass{article}
\usepackage{frascatiphys}
\usepackage{graphicx, amsmath, amssymb, slashed}
\begin{document}
\title{ 
HEAVY NEUTRINOS AT FUTURE COLLIDERS
}
\author{
P. S. Bhupal Dev$^{1,2}$ and  Alejandro Ibarra$^1$       \\[2pt]
{\em $^1$Physik-Department T30d, Technische Univertit\"{a}t M\"{u}nchen,}  \\
{\em James-Franck-Stra\ss e 1, D-85748 Garching, Germany} \\
{\em $^2$Max-Planck-Institut f\"{u}r Kernphysik,} \\
{\em Saupfercheckweg 1, D-69117 Heidelberg, Germany }
}
\maketitle
\baselineskip=11.6pt
\begin{abstract}
We discuss the current status and future prospects of heavy neutrino searches at the energy frontier, which might play an important role in vindicating the simplest seesaw paradigm as the new physics responsible for neutrino mass generation. After summarizing the current search limits and potential improvements at hadron colliders, we highlight the unparalleled sensitivities achievable in the clean environment of future lepton colliders. 
\end{abstract}
\baselineskip=14pt

\section{Introduction}
%The observation of neutrino oscillations in solar, atmospheric, reactor and accelerator neutrino data\cite{Agashe:2014kda} has conclusively established that at least two of the three active neutrinos have a non-zero mass and that individual lepton flavor is violated. This is a concrete experimental evidence for the existence of New Physics beyond the Standard Model (SM). However, 
Despite the spectacular experimental progress in the past two decades in determining the neutrino oscillation parameters, the nature of new physics responsible for non-zero neutrino masses and mixing is still unknown. Given this lack of information, it is essential to explore {\em all} possible ways the neutrino mass mechanism can be probed at various frontiers. In the era of the Large Hadron Collider (LHC), it is therefore natural to ask whether any of the existing theories of neutrino mass can be tested at the energy frontier. %This will be our main focus here.   

A simple paradigm for neutrino masses is the so-called  type-I seesaw mechanism\cite{Minkowski:1977sc},  %,Mohapatra:1979ia,Yanagida:1979as,GellMann:1980vs} 
which postulates the existence of sterile neutrinos ($N$) with Majorana mass $M_N$. Together with the Dirac mass $M_D$, they induce a tree-level active neutrino mass matrix after electroweak symmetry breaking:
\begin{align}
M_\nu \ \simeq \ -M_D M_N^{-1}M_D^T \, .
\label{seesaw}
\end{align}
In a bottom-up phenomenological approach\cite{Drewes:2013gca}, the mass scale of the sterile neutrinos, synonymous with the seesaw scale, is {\em a priori} unknown, and could be anywhere ranging from sub-eV  all the way up to the grand unification theory scale $\sim 10^{15}$ GeV. However, there are arguments based on naturalness of the Standard Model (SM) Higgs mass which suggest the seesaw scale to be  below $\sim 10^7$ GeV\cite{Vissani:1997ys}. Of particular interest to us are TeV-scale seesaw models which are kinematically accessible at the current and foreseeable future collider energies\cite{Deppisch:2015qwa}. Under favorable circumstances, the hadron collider experiments can {\em simultaneously} probe both the key aspects of seesaw, namely, the Majorana nature of the neutrinos and the active-sterile neutrino mixing parameters $V_{\ell N}\equiv M_DM_N^{-1}$ through the ``smoking gun" lepton number violating (LNV) signature of same-sign dilepton plus two jets: $pp\to N\ell^\pm \to \ell^\pm \ell^\pm jj$\cite{Keung:1983uu, Datta:1993nm,Atre:2009rg} and other related processes\cite{Dev:2013wba}. On the other hand, the complementary low-energy probes at the intensity frontier\cite{deGouvea:2013zba, Alekhin:2015byh} are mostly sensitive to only one aspect, e.g. neutrinoless double beta decay ($0\nu\beta\beta$) for the Majorana nature and lepton flavor violation (LFV) searches for the active-sterile neutrino mixing.

\section{Low-scale Seesaw with Large Mixing}

%In the simplest seesaw extension of the SM, i.e. with the minimal addition of the heavy Majorana neutrinos while keeping the SM gauge group intact, there are two key aspects that can be tested experimentally, namely,  the Majorana mass $M_N$ of the mostly sterile neutrinos and their mixing $V_{\ell N}$ with the active neutrinos. 
In the traditional ``vanilla" seesaw mechanism\cite{Minkowski:1977sc}, the active-sterile neutrino mixing parameter  is suppressed by the light neutrino mass $M_\nu \lesssim 0.1~{\rm eV}$:   
\begin{align}
V_{\ell N} \ \simeq \ \sqrt{\frac{M_\nu}{M_N}} \ \lesssim \ 10^{-6} \sqrt{\frac{100~{\rm GeV}}{M_N}}\; .
\label{canon}
\end{align}
Thus for a TeV-scale seesaw, the experimental effects of the light-heavy neutrino mixing are naively expected to be too small, unless the heavy neutrinos have additional interactions, e.g. when they are charged under an additional $U(1)$ or $SU(2)$ gauge group. However, there exists a class of low-scale Type-I seesaw scenarios\cite{Kersten:2007vk, Ibarra:2010xw,Pilaftsis:1991ug}, where $V_{\ell N}$ can be sizable due to specific textures of the Dirac and Majorana mass matrices in Eq.~\eqref{seesaw}. 

Another natural realization of a low-scale seesaw scenario with potentially large light-heavy neutrino mixing is the inverse seesaw mechanism\cite{Mohapatra:1986aw}. In this case, the magnitude of the neutrino mass becomes decoupled from the heavy neutrino mass, thus allowing for a large mixing 
\begin{align}
\label{eq:thetainvseesaw}
		V_{\ell N} \simeq \ \sqrt{\frac{M_\nu}{\mu_S}} \ \approx \ 10^{-2}\sqrt{\frac{1~\text{keV}}{\mu_S}} \; ,
\end{align}
where $\mu_S$ is the only LNV parameter in the theory, whose smallness is `technically natural, i.e. in the limit of $\mu_{S}\to {\bf 0}$,
lepton   number  symmetry   is  restored   and  the light  neutrinos
 are exactly massless  to all  orders in  perturbation theory. 

In the absence of any additional gauge interactions beyond the SM, the amplitudes of the LNV processes in most of the low-scale seesaw models are suppressed by the small mass splitting between the relevant heavy neutrinos, if not by their small mixing with the active neutrino sector, as required to satisfy the light neutrino mass and $0\nu\beta\beta$ constraints\cite{Ibarra:2010xw,Lopez-Pavon:2015cga}.  For collider studies in such situations, one can either use the opposite-sign dilepton signal, relying on the specific kinematic features to separate the signal from the huge SM background, or use the trilepton channel $pp\to N\ell^\pm \to \ell^\pm \ell^\mp \ell^\pm + \slashed{E}_T$\cite{delAguila:2008cj}, which has a relatively smaller cross section, but a smaller SM background as well. In addition, one could look for indirect signatures, such as anomalous Higgs decays induced by the active-sterile neutrino mixing to probe electroweak-scale sterile neutrinos at the LHC\cite{Dev:2012zg}. Introducing new gauge groups beyond the SM and making the sterile neutrinos charged under them enriches the collider phenomenology\cite{Deppisch:2015qwa}, but here our discussion will be limited to the SM seesaw.  
%%%%%%%%%%%%%%%%%%%%%%%%
\section{Searches at Hadron Colliders}

\begin{figure}[t!]
\begin{center}
\includegraphics[width=10cm]{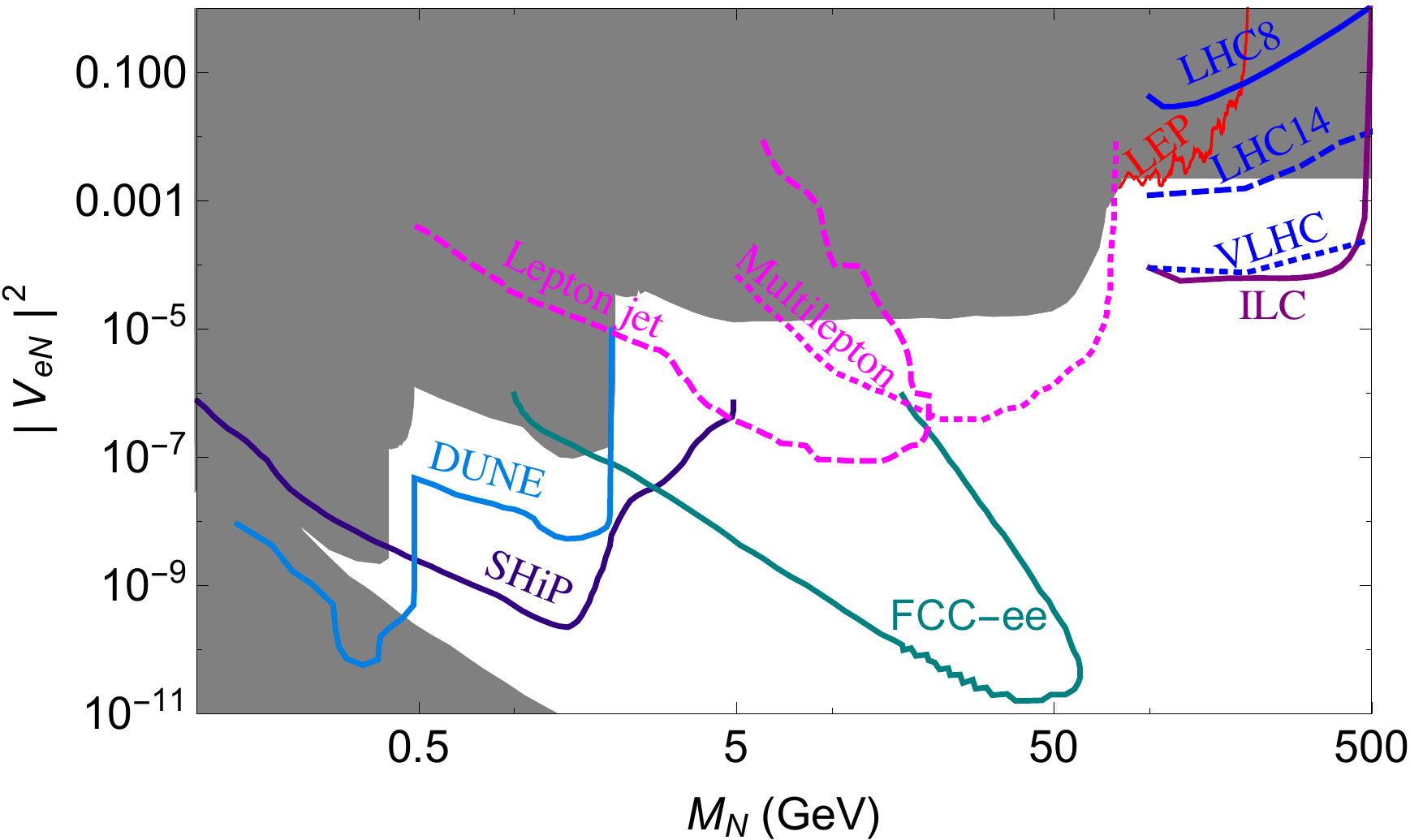}
\end{center}
\caption{Current  (shaded) and future limits in the heavy neutrino mass-mixing plane for the electron flavor. For details and for limits in other flavors, see\cite{Deppisch:2015qwa,Atre:2009rg}.}\label{fig:ven}
\end{figure}

The current direct search limits using the same-sign dilepton channel at $\sqrt s=8$ TeV LHC\cite{Khachatryan:2015gha} range from $|V_{\ell N}|^2 \lesssim 10^{-2}-1$ (with $\ell=e,\mu$) for $M_N=100-500$ GeV. This is shown by the `LHC8' curve in Fig.~\ref{fig:ven} for the electron sector at 95\% confidence level (CL). These limits could be improved by roughly an order of magnitude and extended for heavy neutrino masses up to a TeV or so with the run-II phase of the LHC, as shown by `LHC14' in Fig.~\ref{fig:ven} for 300 fb$^{-1}$ luminosity. Further improvements by another order of magnitude are possible at the proposed 100 TeV $pp$ collider, as shown by the `VLHC' curve for 1 ab$^{-1}$ luminosity. The corresponding limits for opposite-sign dilepton signal are expected to be weaker due to the larger SM background. It is worth emphasizing here that the $W\gamma$ vector boson fusion processes\cite{Dev:2013wba} become increasingly important at higher center-of-mass energies and/or higher masses, and must be taken into account, along with the usual Drell-Yan production mechanism with an $s$-channel $W$ boson so far considered in the experimental analyses of the LHC data.  

Note that the LFV processes (such as $\mu\to e\gamma$) put stringent constraints on the product $|V_{\ell N}V^*_{\ell' N}|$ (with $\ell\neq \ell'$), but do not restrict the individual mixing parameters $|V_{\ell N}|^2$ in a model-independent way. Similarly in the electron sector, the $0\nu\beta\beta$ constraints are the most stringent for a large range of the heavy neutrino masses, but are subject to a large uncertainty due to the unknown $C\!P$ phases in the seesaw matrix, and hence, do not necessarily render the direct searches redundant. The current exclusion limits from various other experiments are shown by the shaded region in Fig.~\ref{fig:ven}\cite{Deppisch:2015qwa,Atre:2009rg}. %The `Seesaw' line indicates the lowest allowed value of the mixing in the seesaw formula, as required by the neutrino oscillation data. 

\section{Searches at Lepton Colliders}
The dominant production channel for heavy neutrinos at $e^+e^-$ colliders is $e^+e^-\to N\nu$ mediated by an $s$-channel $Z$ (for all flavors) and a $t$-channel $W$ (for electron flavor)\cite{Buchmuller:1991tu}. Using the decay channel $N\to eW$ with $W\to jj$, which would lead to a single isolated electron plus hadronic jets, 95\% CL upper limits on $|V_{eN}|^2$ for heavy neutrino mass range between 80 and 205 GeV was derived by LEP\cite{Achard:2001qv}, as shown by the `LEP' contour in Fig.~\ref{fig:ven}. Similar limits were derived\cite{Antusch:2015mia} using the LEP data on $e^+e^-\to W^-W^+\to \bar{\nu}\ell^-\ell^+\nu$. Future lepton colliders can significantly improve the sensitivity in this mass region, as illustrated by the `ILC' curve for $\sqrt s=500$ GeV with 500 fb$^{-1}$ luminosity\cite{Banerjee:2015gca}. Due to its relatively cleaner environment, as compared to hadron colliders, a linear collider thus provides better sensitivity up to heavy neutrino mass values very close to its kinematic threshold. Also note that these limits are valid, irrespective of the Majorana or (pseudo-)Dirac nature of the heavy neutrinos. 

In addition, for heavy Majorana neutrinos, one can explicitly look for LNV processes, such as $e^+e^-\to N e^\pm W^\mp  \to \ell^\pm e^\pm + 4j$\cite{Banerjee:2015gca}. 
Also, switching the beam configuration from $e^+e^-$ to $e^-e^-$ mode, one can also search for the LNV signal $e^-e^-\to W^-W^-\to 4j$ mediated by a $t$-channel Majorana neutrino\cite{Rizzo:1982kn}. 

Before concluding, we should mention that the direct searches discussed above are mostly effective for heavy neutrino masses above 100 GeV or so. For smaller masses, there exist a number of interesting proposals both at energy and intensity frontiers, some of which are shown in Fig.~\ref{fig:ven} labeled as `DUNE'\cite{Adams:2013qkq}, `SHiP'\cite{Alekhin:2015byh}, `FCC-ee'\cite{Blondel:2014bra}, `lepton jet and mulitlepton' at the LHC\cite{Izaguirre:2015pga}. 
%Other indirect searches for sterile neutrinos at colliders include $B$-meson decays at LHCb and anomalous Higgs decays at the LHC.

\section{Conclusion}
Heavy neutrinos are essential constituents of  the simplest seesaw scenario, and hence, their direct searches are important to test the neutrino mass mechanism at the energy frontier. We have briefly reviewed the current status and future prospects of these direct searches for heavy neutrinos at both hadron and lepton colliders. We find that while up to an order of magnitude improvement over the current limit is possible at the LHC, the lepton colliders provide a much better sensitivity due to their clean, almost background-free environment.  
\section{Acknowledgements}
 P.S.B.D. thanks the LFC15 workshop organizers for the invitation and the local hospitality provided by ECT$^*$, Trento. This work was partially supported by the DFG cluster of excellence ``Origin and Structure
of the Universe". P.S.B.D. was also supported in part by a TUM University Foundation Fellowship, and during the completion of this proceedings, by the DFG with grant RO 2516/5-1.

%%%%%%%%%%%%%%%%%%%%%%%%%

\end{document}